\documentclass[manuscript]{aastex}

\slugcomment{Version 2013-05-16}

\shorttitle{Molecular Clouds toward NGC3603}
\shortauthors{Fukui et al.}

\begin{document}
  

\title{Molecular Clouds Toward the Super Star Cluster NGC3603; Possible Evidence for a Cloud-Cloud Collision in Triggering the Cluster Formation}


\author{Y. Fukui\altaffilmark{1},
A. Ohama\altaffilmark{1},
N. Hanaoka\altaffilmark{1},
N. Furukawa\altaffilmark{1, 2},
K. Torii\altaffilmark{1},
J. R. Dawson\altaffilmark{3},
N. Mizuno\altaffilmark{4},
K. Hasegawa\altaffilmark{1},
T. Fukuda\altaffilmark{1},
S. Soga\altaffilmark{1},
N. Moribe\altaffilmark{1},
Y. Kuroda\altaffilmark{1},
T. Hayakawa\altaffilmark{1},
A. Kawamura\altaffilmark{4},
T. Kuwahara\altaffilmark{1},
H. Yamamoto\altaffilmark{1},
T. Okuda\altaffilmark{1, 4},
T. Onishi,\altaffilmark{5},
H. Maezawa\altaffilmark{5},
and A. Mizuno\altaffilmark{6} }
\altaffiltext{1}{Department of Astrophysics, Nagoya University, Chikusa-ku, Nagoya, 464-8602, Japan}
\altaffiltext{2}{present address: Center for Astronomy, Ibaraki University, 2-1-1 Bunkyo, Mito, Ibaraki 310-8512, Japan}
\altaffiltext{3}{School of Mathematics and Physics, University of Tasmania, Sandy Bay Campus, Churchill Avenue, Sandy Bay, TAS 7005}
\altaffiltext{4}{National Astronomical Observatory of Japan, Mitaka, Tokyo, 181-8588, Japan}
\altaffiltext{5}{Department of Astrophysics, Graduate School of Science, Osaka Prefecture University, 1-1 Gakuen-cho, Sakai, Osaka 599-8531, Japan}
\altaffiltext{6}{Solar-terrestrial Environment Laboratory, Nagoya University, Chikusa-ku, Nagoya, 464-8601, Japan}
\email{fukui@a.phys.nagoya-u.ac.jp}



\begin{abstract}
We present new large field observations of molecular clouds with NANTEN2 toward the super star cluster NGC3603 in the transitions  $^{12}$CO($J$=2--1, $J$=1--0) and $^{13}$CO($J$=2--1, $J$=1--0).
We suggest that two molecular clouds at 13\,km\,s$^{-1}$ and 28\,km\,s$^{-1}$ are associated with NGC3603 as evidenced by higher temperatures toward the \ion{H}{2} region as well as morphological correspondence.
The mass of the clouds is too small to gravitationally bind them,  given their relative motion of $\sim$20\,km\,s$^{-1}$.
We suggest that the two clouds collided with each other a Myr ago to trigger the formation of the super star cluster.
This scenario is able to explain the origin of the highest mass stellar population in the cluster which is as young as a Myr and is segregated within the central sub-pc of the cluster.
This is the second super star cluster along side Westerlund2 where formation may have been triggered by a cloud-cloud collision.

\end{abstract}


\keywords{ISM: clouds --- open clusters and associations --- individual: (NGC3603)}



\section{INTRODUCTION}\label{sec:introduction}
High-mass stars dynamically agitate and ionize the interstellar medium in galaxies through strong radiation, stellar winds and, eventually, supernova explosions.
For example, the ultraviolet photons emitted from high-mass stars heat up cold molecular clouds and may inhibit star formation \citep[e.g.,][]{1979MNRAS.186...59W, 1994ApJ...436..795F}.
On the other hand, compressive effects such as the expansion of \ion{H}{2} regions and supernova explosions can collect diffuse gas to enhance star formation \citep[e.g.,][]{1977ApJ...214..725E, 2006ApJ...648L.131H}.
These feedback effects play an important role in regulating star formation in galaxies.

High-mass stars must be formed rapidly in dense molecular clumps  in order to overcome the feedback effects from forming stars such as radiation pressure which inhibits mass accretion \citep{2004ApJ...603..383T}.
The alternotive scenario of stellar merging requires unreasonably large stellar densities of $\sim 10^{8}$\,Mpc$^{-3}$ \citep{1998MNRAS.298...93B} and is not feasible in high-mass star formation in most galaxies.
\citet{1998ASPC..148..150E} outline three mechanisms by which the formation and collapse of dense clumps in giant molecular clouds (GMCs) may be dynamically triggered by external events. 
The first is globule squeezing, in which pre-existing dense clumps are compressed, either by high ambient pressures in \ion{H}{2} regions, or by shock-waves propagating from supernovae or other disturbances.
The second is ``collect and collapse'', in which gas accumulated into a shell or ridge by expanding \ion{H}{2} regions, stellar winds or supernovae collapses to form new dense clumps and stars.
The third is cloud-cloud collisions, where two molecular clouds collide leading to star formation via gravitational instabilities.

Most high-mass stars are formed in stellar clusters.
Super star clusters (SSCs) are the most massive clusters in the Galaxy, with stellar densities exceeding $10^4$ stars\,pc$^{-3}$ in their cores \citep{johnson05}.
There are several known SSCs in external galaxies.
For example, R136 in the Large Magellanic Cloud \citep{1995ApJ...444..758H, 1998ApJ...493..180M} and NGC 604 in M33 \citep{1995ApJS...99..551H, 2004AJ....128.1196M}.

In addition, the two colliding Antennae galaxies NGC4038 and NGC4039 have young massive star clusters of $10^{5}$--$10^{6}$\,$M_{\odot}$, distributed across several kpc \citep{2000ApJ...542..120W}.
Collisions between galaxies will accumulate the ISM in a compact region leading to the formation of gravitationally unstable dense clumps, and it is likely that such a collision can trigger massive cluster formation. 
It is however difficult to resolve the distribution of the parent molecular clouds in the Antennae galaxies because of their large distance, 20\,Mpc.
The detailed mechanism of SSC formation is therefore still a mystery because detailed observations of the parent clouds are not available at a sufficiently high spatial resolution. 
Within the Milky Way there are 8 known young SSCs with an age of a few Myrs \citep{2010ARA&A..48..431P}.
They are the Arches, Quintuplet, RCW38, Westerlund 2, Tr14, NGC3603, Westerlund 1 and [DBS2003]179; in four of them (the Arches, Quintuplet, Tr14 and Westerlund 1) their parent molecular clouds appear to have been dissipated already due to stellar winds and/or ultraviolet photons (UV)  (see Table \ref{tbl:ssc_properties}).
In the remaining four SSCs (RCW38, Westerlund 2,  NGC3603 and [DBS2003]179) we see signs of associated nebulosity  at infrared wavelengths, and they are good candidates for identifying parent clouds. 

\citet{furukawa09} and \citet{2010ApJ...709..975O} presented observational evidence that a cloud-cloud collision may have triggered the formation of the SSC Westerlund 2.
These authors report two GMCs centered at 4\,km\,s$^{-1}$ and 16\,km\,s$^{-1}$. 
They argue that acceleration by stellar winds from Westerlund 2 is insufficient to explain the entire observed velocity dispersion of the molecular gas, and suggest a scenario in which a collision between the two clouds may have triggered the formation of the SSC. 
Theoretical studies have modelled the triggering of star formation via cloud-cloud collisions \citep{1992PASJ...44..203H, 2010MNRAS.405.1431A}, and such collisions are believed to be frequent in gas-rich galaxies \citep{2009ApJ...700..358T}.

Following the molecular line studies on Westerlund 2, this paper focuses on NGC 3603 which is similar in stellar age and mass. 
NGC 3603 is located at ($l$, $b$)=(291{\fdg}6, $-$0{\fdg}5), close to the tangent of the Carina spiral arm. 
Discovered by Sir John Herschel in 1834, it is known at present as one of the most massive clusters in the Milky Way \citep{1969ApL.....4..199G, 1973ApJ...182L..21W, 1983A&A...124..273M}.
The estimated age of the cluster is 1 -- 2\,Myr \citep[e.g.,][]{2004AJ....127.1014S, 2008ApJ...675.1319H} and the total stellar mass has been estimated to be 1.0 -- $1.6 \times 10^4$\,$M_{\sun}$ \citep{2008ApJ...675.1319H} and $1.76 \pm 0.38 \times 10^4$\,$M_{\sun}$ \citep{2010ApJ...716L..90R}.
This rich stellar population includes more than 30 O type stars \citep{2004AJ....128.2854M} and several Wolf-Rayet (WR) stars \citep{1999RMxAC...8...41S} producing a Lyman continuum flux of $1.5 \times 10^{39}$\,erg\,s$^{-1}$ \citep{1984ApJ...287..116K, 1995AJ....110.2235D}, corresponding to about 100 times the ionizing power of the Orion Trapezium cluster.
NGC 3603 is therefore an ideal experimental laboratory in which to study a galactic prototype of the starburst clusters that are observed in the other galaxies. 
The distance of NGC 3606 has been estimated to be 6.1 -- 6.3\,kpc \citep{1969ApL.....4..199G, 1999AJ....117.2902D, 2000PASJ...52..847P, 2004AJ....128..765S}, 7.0 -- 7.3\,kpc \citep{1983A&A...124..273M, 1978A&A....63..275V, 1987A&A...171..261C, 1989A&A...213...89M} and as high as 8 or even 10\,kpc \citep{1969ApL.....4..199G, 1974A&A....35..315M, 1998MNRAS.296..622C}. 
In the present work we shall adopt a distance of 7\,kpc, the mean of the published values.

Only a few molecular line observations of the NGC 3603 region have been reported to date.
\citet{1988ApJ...331..181G} carried out a large-scale $^{12}$CO($J$=1--0) observations of the Carina region at a 8{\farcm}8 resolution and identify a molecular cloud at 15 km s$^{-1}$ toward NGC3603 listed as cloud No.\ 17.
The total molecular mass of the cloud is $4 \times 10^{5}$\,$M_{\sun}$.
These authors also identified another cloud near NGC3603 at 29\,km\,s$^{-1}$ (No.\ 18), with an estimated molecular mass of $1 \times 10{^5}$\,$M_{\sun}$.
CS observations by \citet{2002A&A...394..253N} showed that there are 13 massive molecular clumps around 15\,km\,s$^{-1}$ in the NGC3603 region.
Most recently, \citet{2011A&A...525A...8R} used NANTEN2 to make CO($J$=4--3) observations of NGC3603 for an area of $2{\arcmin} \times 2{\arcmin}$ in $l$ and $b$ and revealed that two molecular clumps at 13\,km\,s$^{-1}$, MM1 and MM2, are associated with two pillars with photo dissociation region (PDRs) observed by the HST \citep{2000AJ....119..292B}.
NGC 3603 ionizes an extensive and luminous \ion{H}{2} region, RCW57B, with the Wolf-Rayet star WR HD97950 at its center \citep{1995A&A...300..403H}.
The intensive radiation and stellar winds from the cluster shape large gaseous pillars \citep{1999A&A...352L..69B}  at the edge of the cloud \citep{2000AJ....119..292B}.
There is also another \ion{H}{2} region, NGC3576, located 0.4\,degrees to the southwest of NGC3603, and its associated molecular cloud is listed as No.\ 20 in \citet{1988ApJ...331..181G}.
However, NGC3576 is on the near side of the Carina arm at a distance of 3.0\,kpc and is not related to NGC3603 \citep{1999AJ....117.2902D}.

In this paper, we present new CO($J$=2--1) and ($J$=1--0) observations of NGC3603 with NANTEN2, and an analysis of this molecular data set.
Section \ref{sec:observations} describes the observations.
Section \ref{sec:results} presents the molecular data and in Section \ref{sec:discussion} we discuss the formation of the super star cluster.
Conclusions are summarized in Section \ref{sec:summary}.

\section{OBSERVATIONS}\label{sec:observations}
Observations of the $J$=2--1 transition of CO were made with the NANTEN2 4\,m sub-millimeter telescope of Nagoya University at Atacama (4865\,m above sea level) in October-November  2008 for $^{12}$CO($J$=2--1) and in October 2009 for $^{13}$CO($J$=2--1).
The half-power beam width (HPBW) of the telescope was 90{\arcsec} at 230\,GHz.
The 4\,K cooled superconductor-insulator-superconductor (SIS) mixer receiver provided a typical system temperature of $\sim$200\,K in a single-side band at 220 -- 230\,GHz, including the atmosphere toward the zenith.
The spectrometer was an acousto-optical spectrometer (AOS) with 2048 channels, providing velocity coverage of 392\,km\,s$^{-1}$ at 230\,GHz.
The pointing was checked regularly by observing the radio continuum emission from Jupiter, and was accurate to within 10{\arcsec}.
The target region was observed between elevation angles of 30{\arcdeg} and 60{\arcdeg}.
We observed a large area surrounding NGC 3603 in $^{12}$CO($J$=2--1), while $^{13}$CO($J$=2--1) observations are limited to a smaller region (see Figure \ref{fig:overview}).
The OTF (on-the-fly) mapping mode was used in the observations, and the output grid of the region is 30{\arcsec}.
We smoothed the velocity and spatial resolutions to 0.19\,km\,s$^{-1}$ and 100{\arcsec}, respectively, to achieve a better noise level.
Finally, we obtained rms noise fluctuations of $\sim$0.2\,K and $\sim$0.1\,K per channel in $^{12}$CO($J$=2--1) and $^{13}$CO($J$=2--1), respectively.
The standard sources Ori KL ($\alpha$, $\delta$)$_\mathrm{J2000}$=(5$^\mathrm{h}$~32$^\mathrm{m}$~14{\fs}5, $-$5{\arcdeg}~22{\arcmin}~27{\farcs}6) for $^{12}$CO($J$=2--1) and  $\rho$ Oph East ($\alpha$, $\delta$)$_\mathrm{J2000}$=(16$^\mathrm{h}$~32$^\mathrm{m}$~22{\fs}56, $-$24{\arcdeg}~28{\arcmin}~31{\farcs}8) were observed for intensity calibration  for $^{13}$CO($J$=2--1) every 2 hours.
We assumed true main beam temperatures, $T_\mathrm{MB}$, of 75 -- 83\,K in $^{12}$CO($J$=2--1) as observed by the KOSMA telescope \citep{1998A&A...335.1049S} and 17\,K in $^{13}$CO(J = 2 --1) as observed by the 60\,cm Survey telescope \citep{2007PASJ...59.1005N}.

Observations of the $^{12}$CO($J$=1--0) and $^{13}$CO($J$=1--0) transitions were made with NANTEN2 telescope during  September to November 2011.
The observations were carried out with a 4\,K cryogenically cooled Nb SIS mixer receiver.
The typical system temperature was $\sim$270\,K in the double-side band.
Two digital spectrometers provided a bandwidth and resolution of 1\,GHz and 61\,kHz, which corresponds to 2600\,km\,s$^{-1}$ with velocity resolution of 0.17\,km s$^{-1}$, respectively, at 110\,GHz.
The pointing was checked regularly on the Sun by radio continuum emission.
The HPBW of the telescope was 2{\farcm}6.
Observations of $^{12}$CO($J$=1--0) and $^{13}$CO($J$=1--0) were simultaneously made in the OTF mode with a 1{\arcmin} grid spacing.
We smoothed the velocity and spatial resolutions to 0.66\,km\,s$^{-1}$ and 163{\arcsec}, respectively.
Finally, we obtained rms noise fluctuations of $\sim$0.3\,K and $\sim$0.2\,K per channel in $^{12}$CO($J$=1--0) and $^{13}$CO($J$= 1--0), respectively.
We used the standard sources Ori KL ($\alpha$, $\delta$)$_\mathrm{J2000}$=(5$^\mathrm{h}$~32$^\mathrm{m}$~14{\fs}5, $-$5{\arcdeg}~22{\arcmin}~27{\farcs}6) for $^{12}$CO($J$=1--0)  and $^{13}$CO($J$=1--0).

\section{RESULTS}\label{sec:results}
\subsection{Morphological and Kinematic Analysis}\label{subsec:morphology}
\begin{figure}
\epsscale{1}
\plotone{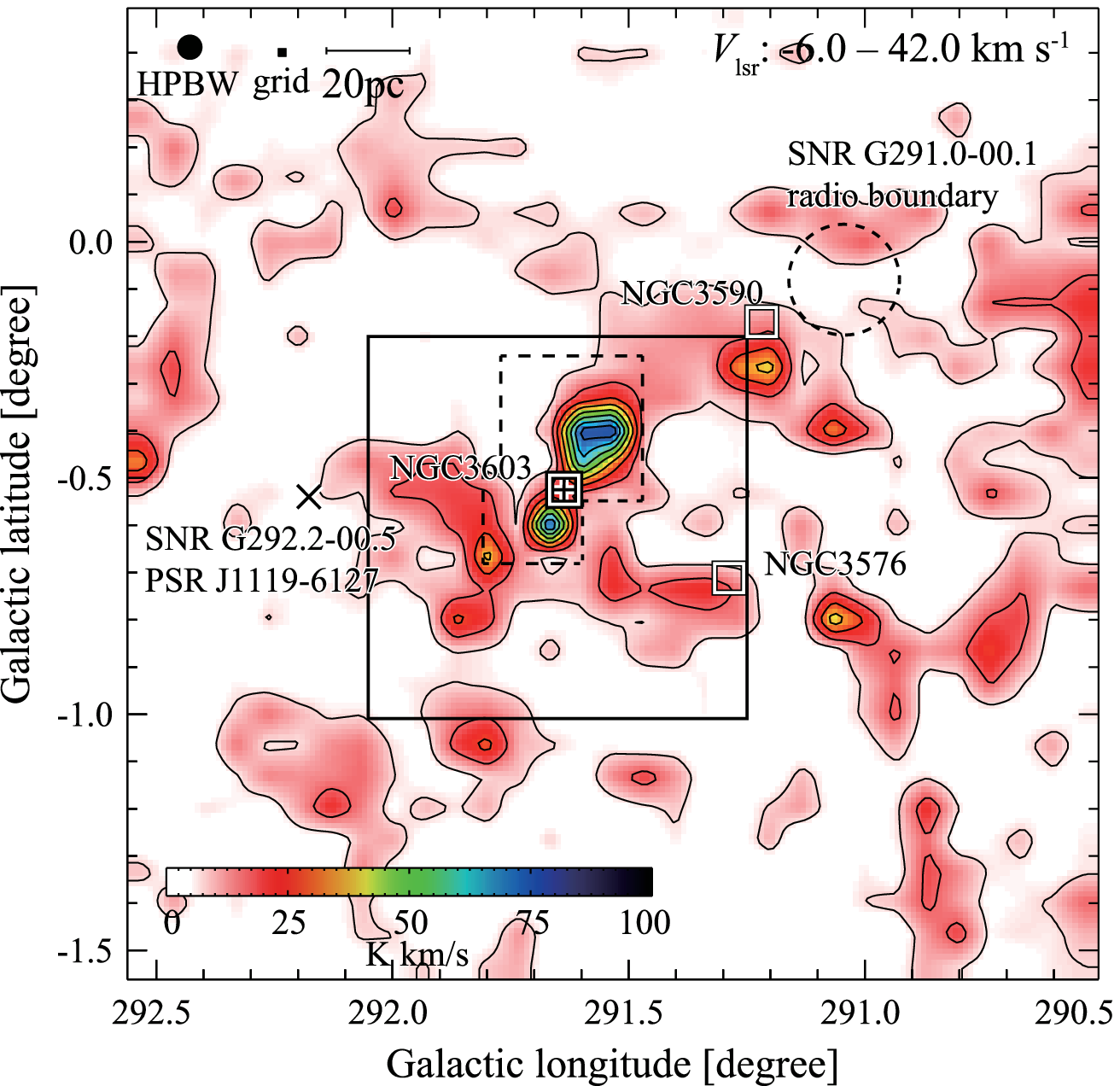}
\caption{Integrated intensity distribution of NANTEN2  $^{12}$CO($J$=1--0) emission toward ($l$, $b$) = (290{\fdg}5 -- 292{\fdg}5, $-$1{\fdg}5 -- 0{\fdg}5).
Contours are drawn every 10\,K\,km\,s$^{-1}$ from 5\,K\,km\,s$^{-1}$.
The square indicates NGC 3603, and solid lines indicate the region shown in Figures \ref{fig:chmap_CO21} -- \ref{fig:chmap_13CO10}.
The dashed lines show the region observed in $^{13}$CO($J$=2-1).}
\label{fig:overview}
\end{figure}

\begin{figure}
\epsscale{1}
\plotone{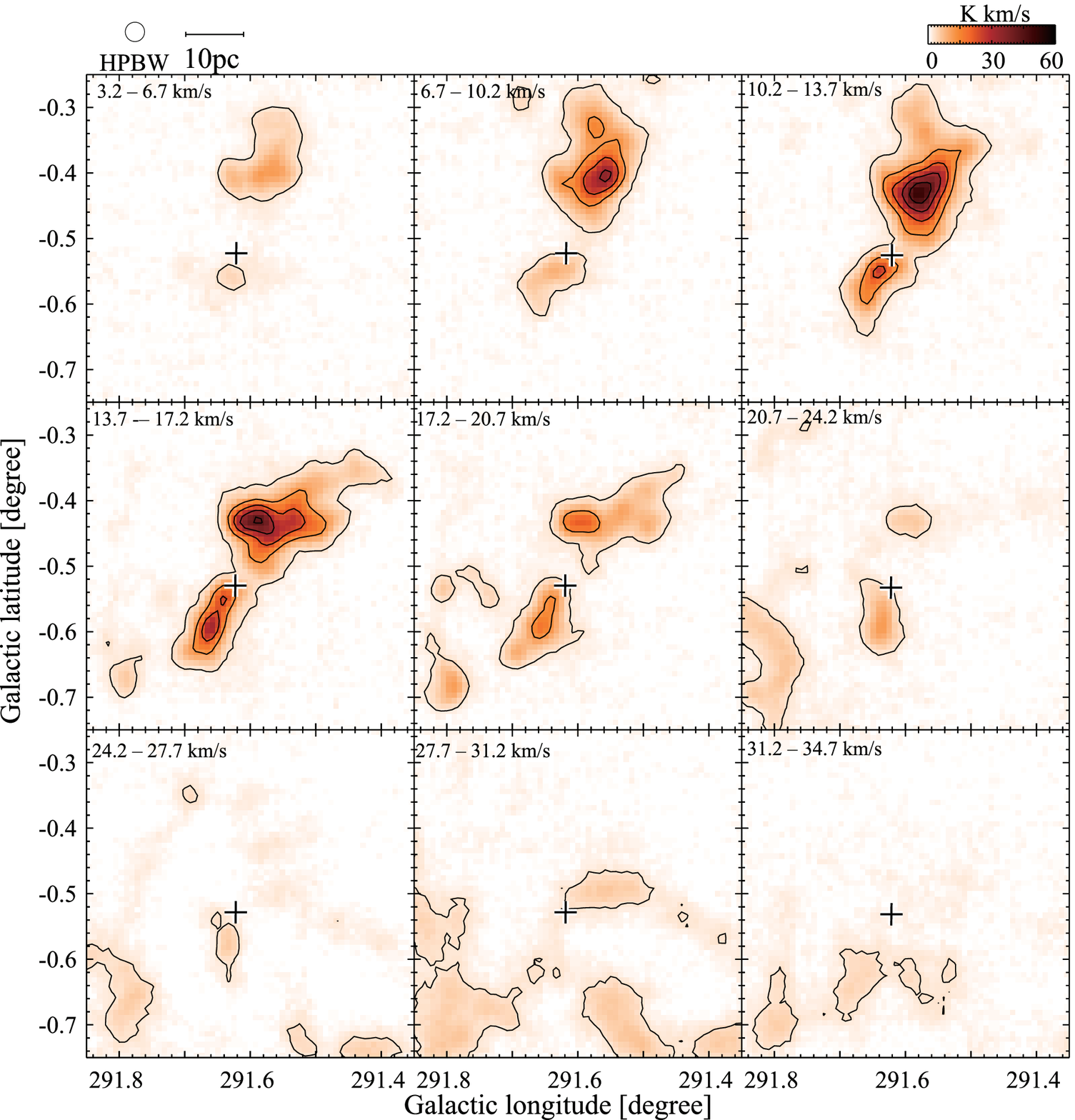}
\caption{Velocity channel maps of $^{12}$CO($J$=2--1) intensity integrated over 3.5\,km\,s$^{-1}$ bins.
Contours are drawn every 6.0\,K\,km\,s$^{-1}$ from 2.4\,K\,km\,s$^{-1}$.
The cross corresponds to the position of the cluster NGC3603.}
\label{fig:chmap_CO21}
\end{figure}

\begin{figure}
\epsscale{0.8}
\plotone{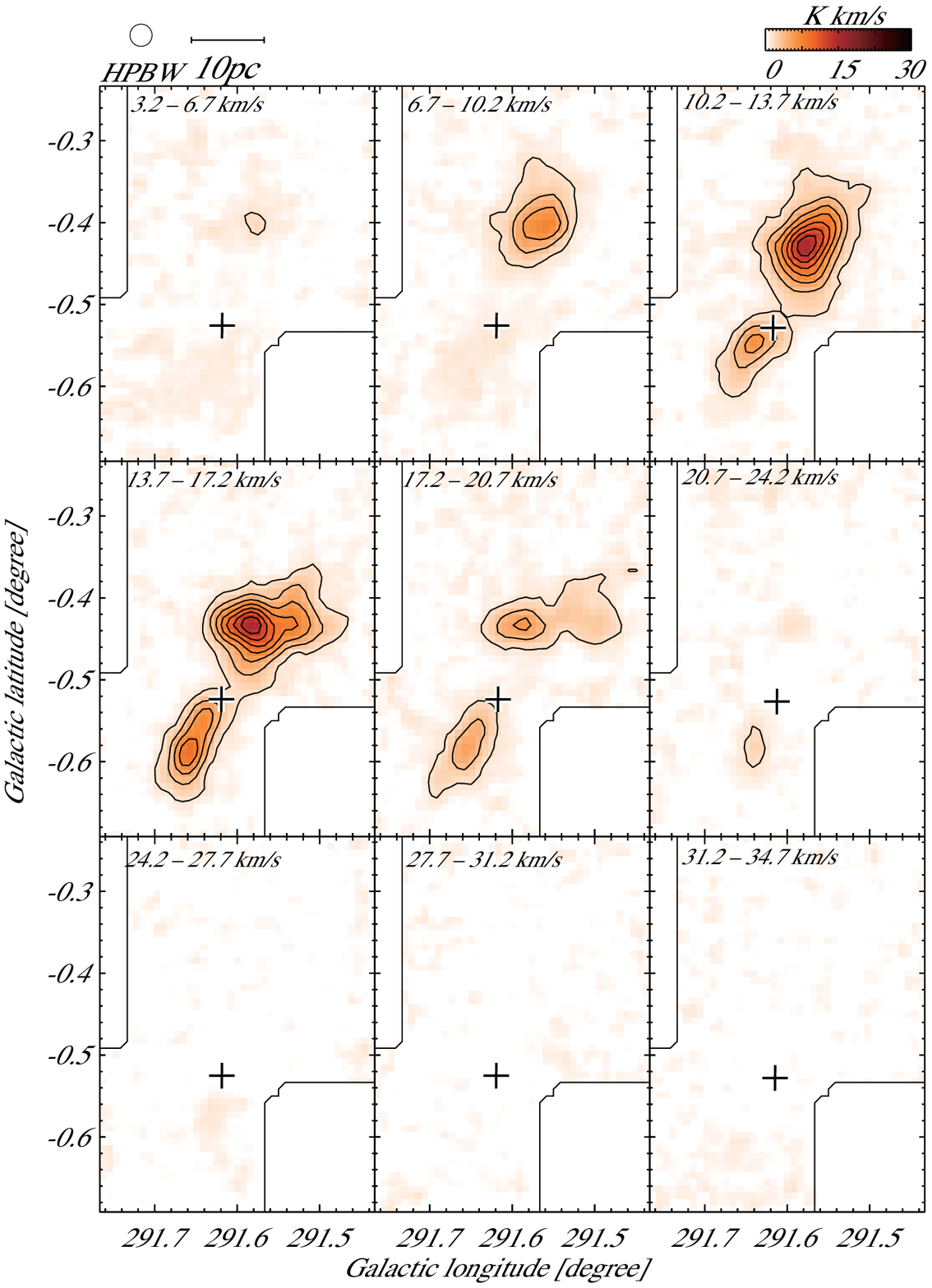}
\caption{Velocity channel maps of $^{13}$CO($J$=2--1) intensity integrated over 3.5\,km\,s$^{-1}$ bins.
Contours are drawn every 2.0\,K\,km\,s$^{-1}$ from 1.3\,K\,km\,s$^{-1}$.
The cross corresponds to the position of the cluster NGC3603.}
\label{fig:chmap_13CO21}
\end{figure}

\begin{figure}
\epsscale{1}
\plotone{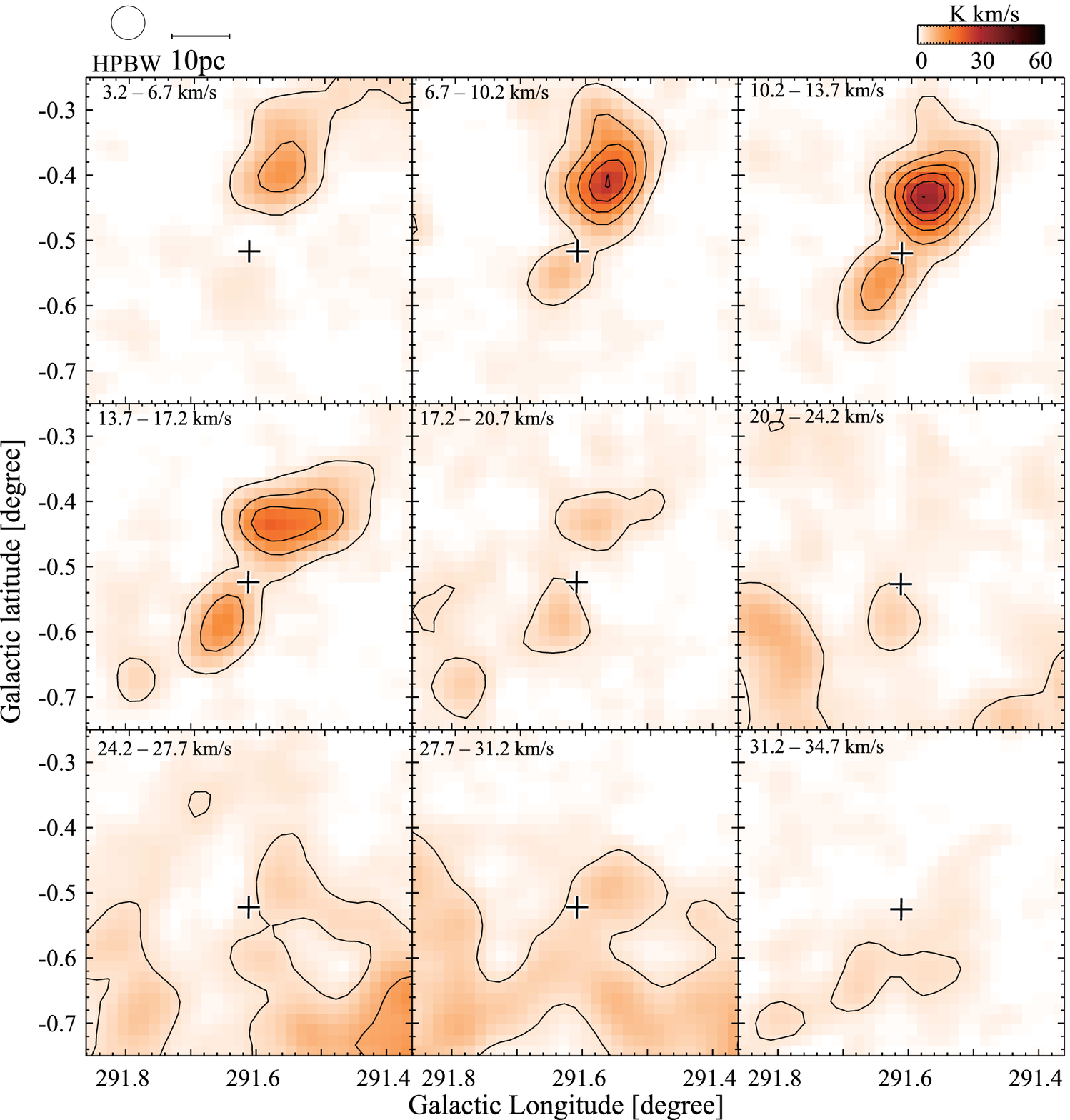}
\caption{Velocity channel maps of $^{12}$CO($J$=1--0) intensity integrated over 3.5\,km s$^{-1}$ bins.
Contours are drawn every 6.0\,K\,km\,s$^{-1}$ from 2.4\,K\,km\,s$^{-1}$.
The cross corresponds to the position of the cluster NGC3603.}
\label{fig:chmap_CO10}
\end{figure}

\begin{figure}
\epsscale{1}
\plotone{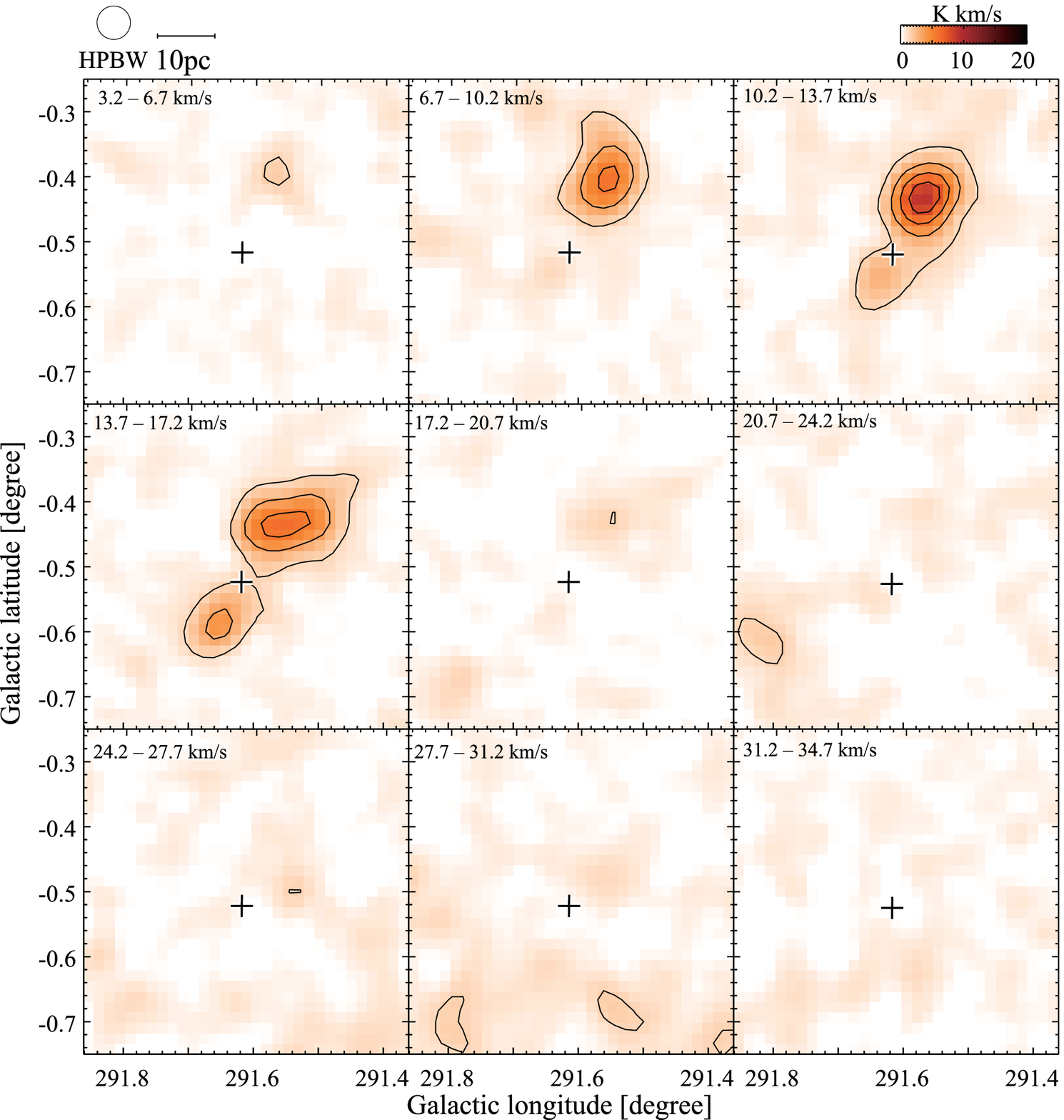}
\caption{Velocity channel maps of $^{13}$CO($J$=1--0) intensity integrated over 3.5\,km\,s$^{-1}$ bins.
Contours are drawn every 2.0\,K\,km\,s$^{-1}$ from 1.4\,K\,km\,s$^{-1}$.
The cross corresponds to the position of the cluster NGC3603.}
\label{fig:chmap_13CO10}
\end{figure}

Figure \ref{fig:overview} shows the large scale $^{12}$CO($J$=1--0) distribution toward NGC3603, including only the positive velocity clouds on the far side of the Carina arm \citep{mizuno04}.
NGC3603 is located between the two brightest peaks of $^{12}$CO($J$=1--0) in the center of Figure \ref{fig:overview}.
Figures \ref{fig:chmap_CO21} -- \ref{fig:chmap_13CO10} show the velocity channel distributions of the $^{12}$CO($J$=2--1), $^{13}$CO($J$=2--1), $^{12}$CO($J$=1--0) and $^{13}$CO($J$=1--0) emission, respectively.
The velocity range of these figures (from 3.2\,km\,s$^{-1}$ to 34.7\,km\,s$^{-1}$) was chosen to include all features in the vicinity of NGC3603.
The bright CO cloud peaked at around 13\,km\,s$^{-1}$ \citep[cloud No.\ 17,][]{1988ApJ...331..181G} has two peaks on the north and south of NGC3603, and is obviously associated with NGC3603 and the other weak CO features at around 28\,km\,s$^{-1}$ are additional candidates for associated clouds \citep[cloud No.\ 18,][]{1988ApJ...331..181G}.
We name the former the blue-shifted cloud and the latter the red-shifted cloud.
The blue-shifted cloud is compact and intense, while the red-shifted cloud is extended and weak.
We give the parameters toward the two peaks of the two clouds in Table \ref{tbl:cloud_properties}.
The molecular column density is generally ten-times higher in the blue-shifted cloud than in the red-shifted cloud.

\begin{figure}
\epsscale{1}
\plotone{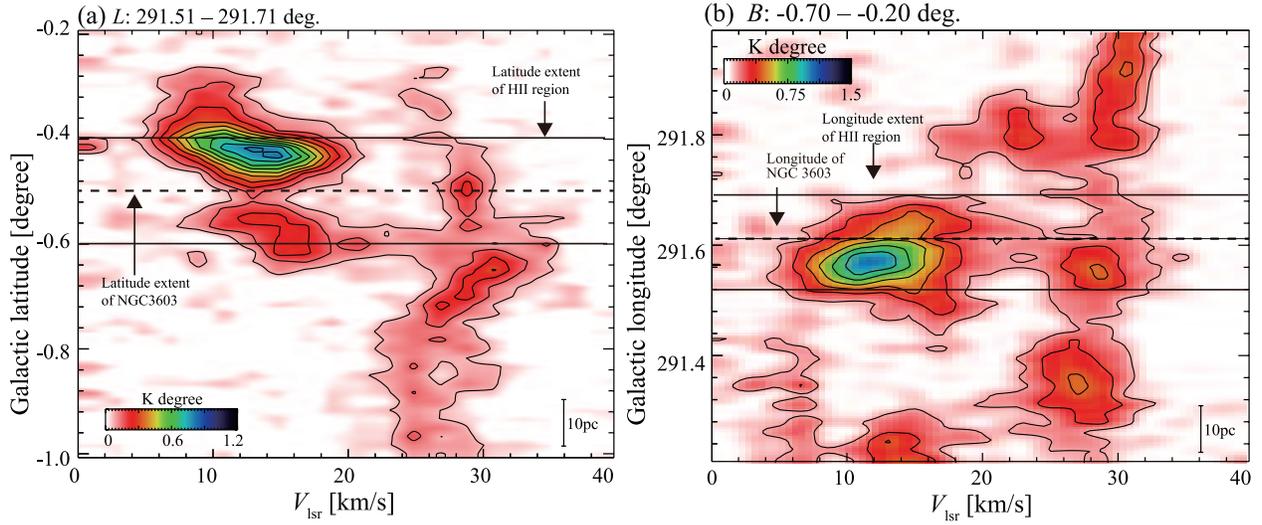}
\caption{(a) Velocity vs.\ galactic latitude diagram for $^{12}$CO($J$=1--0) emission integrated over a longitude range of 291{\fdg}51 -- 291{\fdg}71.
Contours are plotted every 0.06\,K\,degree from 0.02\,K\,degree.
The solid lines depict the extent of the \ion{H}{2} region observed by the \textit{Spitzer Space Telescope}.
The dashed line depicts the latitude position of NGC3603.
(b) Velocity vs.\ galactic longitude diagram for $^{12}$CO($J$=1--0) emission integrated over a latitude range of $-$0{\fdg}70 -- $-$0{\fdg}20.
Contours are plotted every 0.15\,K\,degree from 0.15\,K\,degree.}
\label{fig:pvdiagram}
\end{figure}

Figures \ref{fig:pvdiagram} a and b show $^{12}$CO($J$=1--0) position-velocity diagrams of NGC3603.
We see the two velocity components at 13\,km\,s$^{-1}$ and 28\,km\,s$^{-1}$ and a bridging feature in the velocity range 18 -- 25\,km\,s$^{-1}$ between the two clouds at $b=-0{\fdg}6$ in Figure \ref{fig:pvdiagram} a and at $l=291{\fdg}5$ -- 291{\fdg}7 in Figure \ref{fig:pvdiagram} b.

\begin{figure}
\epsscale{1}
\plotone{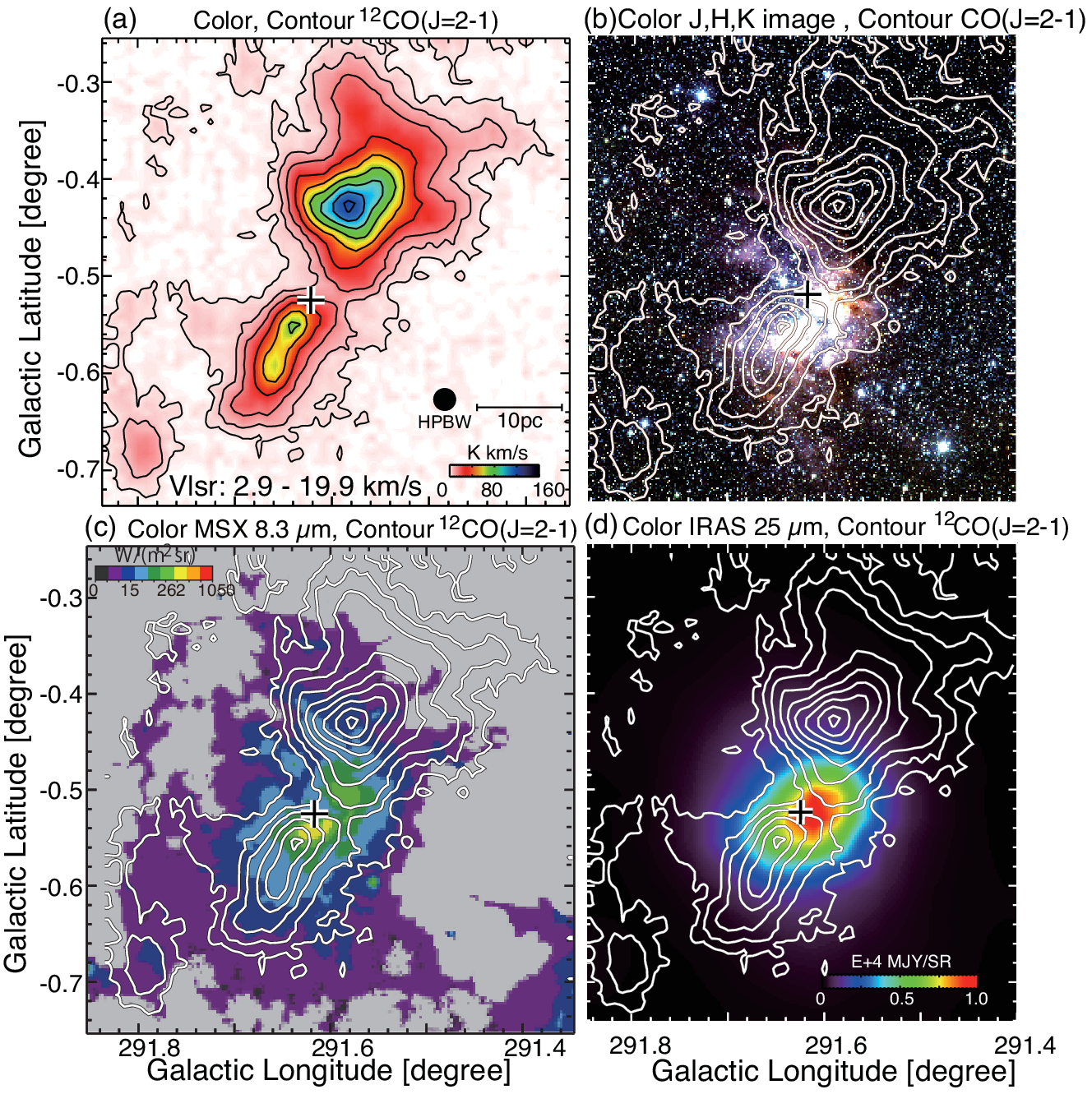}
\caption{(a) Distribution of $^{12}$CO($J$=2--1) integrated intensity with the velocity range from 2.9\,km\,s$^{-1}$ to 19.9\,km\,s$^{-1}$.
Contours are plotted at levels of (2.5, 7.0, 15, 30, 50, 70, 90)\,K\,km\,s$^{-1}$.
A cross depicts the central cluster.
(b) $^{12}$CO($J$=2--1) distribution of the molecular clouds associated with NGC3603, superposed on $J\!H\!K$ image \citep{2003yCat.2246....0C}.
Blue, yellow and red show $J$, $H$, and $K$ bands, respectively.
(c) Contour maps of the $^{12}$CO($J$= 2--1) emission superposed on the \textit{MSX} 8.3\,{\micron} image \citep{2010ScChG..53S.271W}.
The cross depicts the central cluster.
(d) Contour maps of the $^{12}$CO($J$=2--1) emission superposed on IRAS 25\,{\micron} data.}
\label{fig:blueshifted_cloud}
\end{figure}

\begin{figure}
\epsscale{1}
\plotone{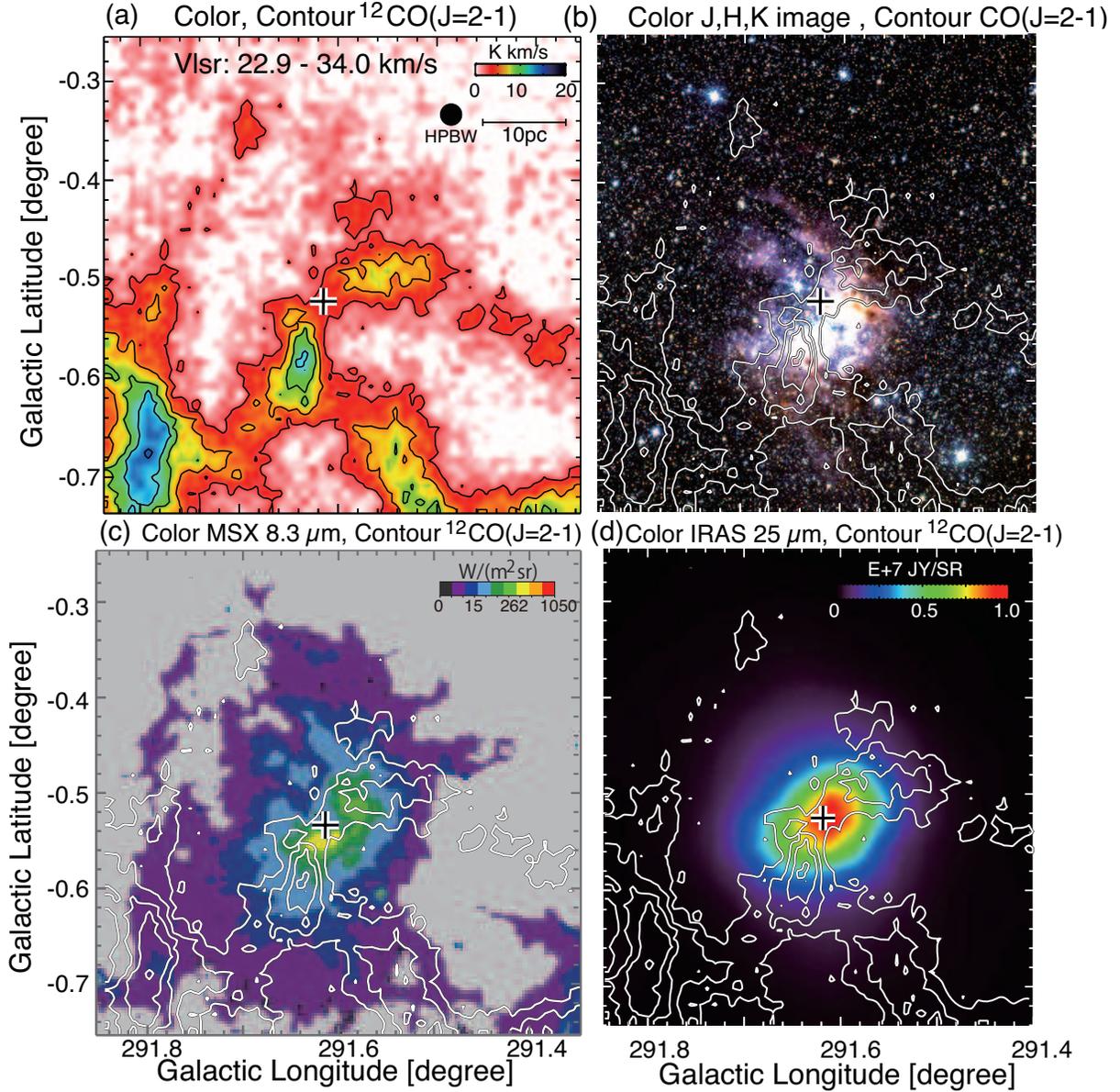}
\caption{(a) Distribution of $^{12}$CO($J$=2--1) integrated intensity with the velocity range from 22.9\,km\,s$^{-1}$ to 34.0\,km\,s$^{-1}$.
Contours are plotted every 3.0\,K\,km\,s$^{-1}$ from 3.0\,K\,km\,s$^{-1}$.
A cross depicts the central cluster.
(b) $^{12}$CO($J$=2--1) distribution of the molecular clouds associated with NGC3603,  superposed on $J\!H\!K$ image \citep{2003yCat.2246....0C}.
Blue, yellow and red show $J$, $H$, and $K$ bands, respectively.
(c) Contour maps of the $^{12}$CO($J$=2--1) emission superposed on the \textit{MSX} 8.3\,{\micron} image \citep{2010ScChG..53S.271W}.
The cross depicts the central cluster.
(d) Contour maps of the $^{12}$CO($J$=2--1) emission superposed on IRAS 25\,{\micron} data.}
\label{fig:redshifted_cloud}
\end{figure}

Figures \ref{fig:blueshifted_cloud} and \ref{fig:redshifted_cloud} show the distributions of the two clouds in $^{12}$CO($J$=2--1) overlaid on the infrared images in $J\!H\!K$ bands, 8.3\,{\micron}, and 25\,{\micron}.
For the blue-shifted cloud, we see the molecular distribution is correlated with the cluster and the \ion{H}{2} region.
First, the cloud shows a depression toward the cluster at all CO lines, which suggests that the molecular gas toward the cluster has been dispersed by ionization/stellar winds.
Second, the $J\!H\!K$ distribution (Figure \ref{fig:blueshifted_cloud} b) shows that the northern peak of the blue-shifted cloud coincides with K band obscuration whose southeastern edge is delineated clearly at ($l$, $b$ = 291{\fdg}5 -- 291{\fdg}65, $-$0{\fdg}5 -- $-$0{\fdg}55) (Figures \ref{fig:blueshifted_cloud} c and d).
Obscuration by the red-shifted cloud is too small to affect the near infrared image ($A_{J}$, $A_{H}$, $A_{K}=0.1$ -- 0.7\,mag, see Table \ref{tbl:cloud_properties}).
It is unclear the cloud is in front or behind of the cluster, but we also see a hint of association between the cloud and the cluster/\ion{H}{2} region; the cloud shows a similar depression toward the cluster, suggesting cloud dispersal due to the cluster.

\subsection{Temperature and Density of the Molecular Clouds}
In order to investigate the temperature of the molecular gas, which is a good indicator of physical association of the clouds with the cluster, we first examine the ratio of the CO($J$=2--1) and CO($J$= 1--0) line intensities.

\begin{figure}
\epsscale{0.8}
\plotone{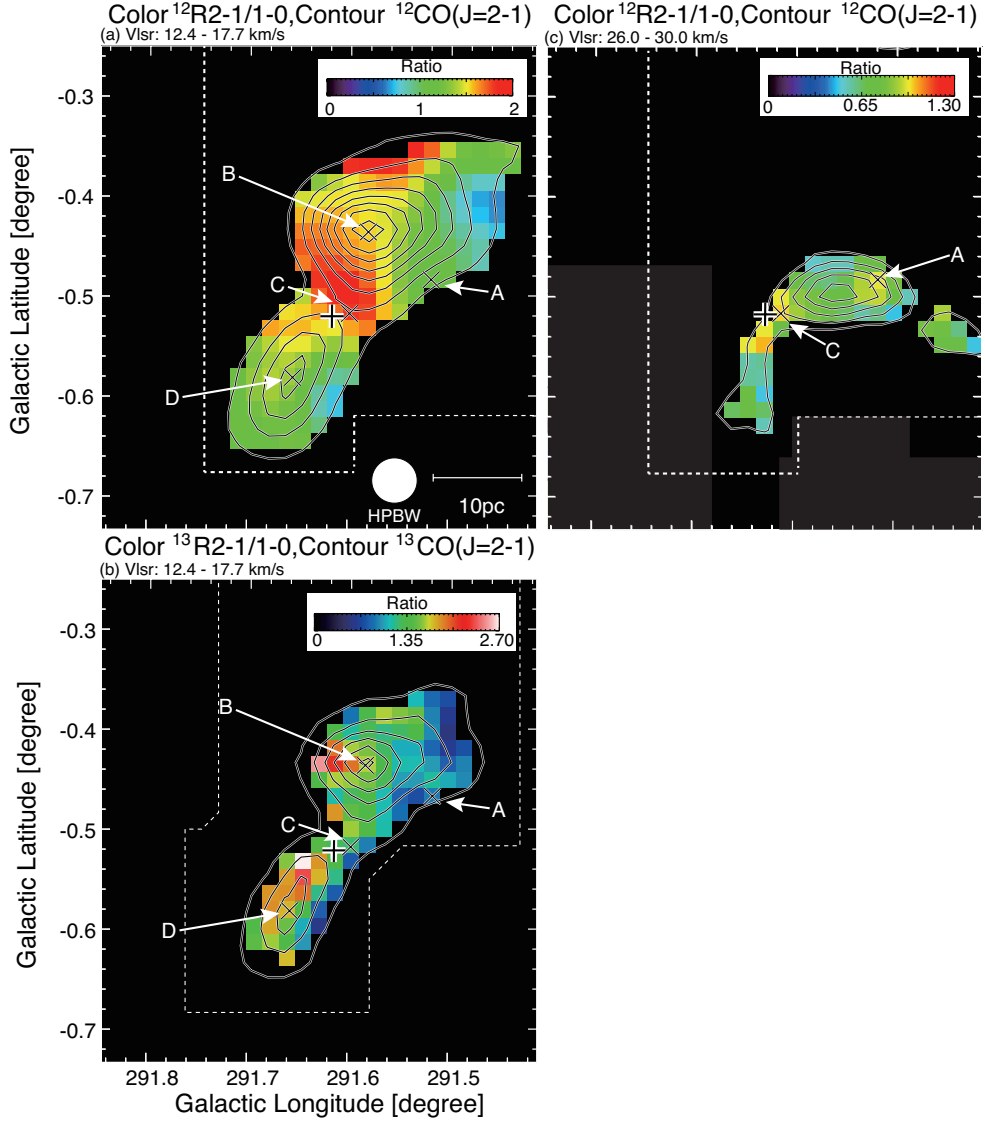}
\caption{(a) Distribution of the ratio of $^{12}$CO($J$=2--1) and $^{12}$CO($J$=1--0) integrated intensity in the range 12.4\,km\,s$^{-1}$ to 17.7\,km\,s$^{-1}$.
Contours show the $^{12}$CO($J$=2--1) emission smoothed with a Gaussian function to a 156{\arcsec} spatial resolution and are plotted every 4.5\,K\,km\,s$^{-1}$ from 4.5\,K\,km\,s$^{-1}$.
(b) Distribution of the ratio of $^{13}$CO($J$=2--1) and $^{13}$CO($J$=1--0) integrated intensity in the range 1.0\,km\,s$^{-1}$ to 2.5\,km\,s$^{-1}$.
(c) Distribution of the ratio of $^{12}$CO($J$=2--1) and $^{12}$CO($J$=1--0) integrated intensity in the range 27.6\,km\,s$^{-1}$ to 34.3\,km\,s$^{-1}$.
The cross corresponds to the position of the cluster NGC3603.
Points A--D show the locations of the CO spectra in Figure \ref{fig:CO_spectra} and LVG analysis in Figure \ref{fig:LVG1}.}
\label{fig:intensity_ratio}
\end{figure}

Figures \ref{fig:intensity_ratio} a and b show the distributions of the ratio of $J$=2--1 to $J$=1--0 line integrated intensities in $^{12}$CO and $^{13}$CO, for the blue-shifted cloud.
The $^{12}$CO distribution shows that the ratio is enhanced significantly near the cluster.
We also see another enhancement of the ratio at the northern edge of the cloud.
The high ratio of $^{12}$CO (above 1.0) suggests high temperatures due to extra heating by high-mass stars, since the typical ratio is around 0.6 in clouds with no extra heat source \citep[e.g.,][]{1997ApJ...486..276S, 2011ApJ...738...46T}.
We infer that the blue-shifted cloud as a whole is heated-up by the cluster.
The heating is especially significant in the region within $\sim$5\,pc of the cluster, where the ratio is higher than 1.5.
The $^{13}$CO ratio is also enhanced to 1.0 -- 2.7 within 10\,pc of the cluster.
We suggest that the irradiated surface layer of the cloud is better traced in the optically thick $^{12}$CO than in the optically thin $^{13}$CO.
The red-shifted cloud is not significantly detected in $^{13}$CO and we show only the distribution of $^{12}$CO ratio in Figure \ref{fig:intensity_ratio} c.
The red-shifted cloud also shows enhanced ratios above 1.0 within a few pc of the cluster, suggesting the cloud is also heated-up by the cluster.

\begin{figure}
\epsscale{1.}
\plotone{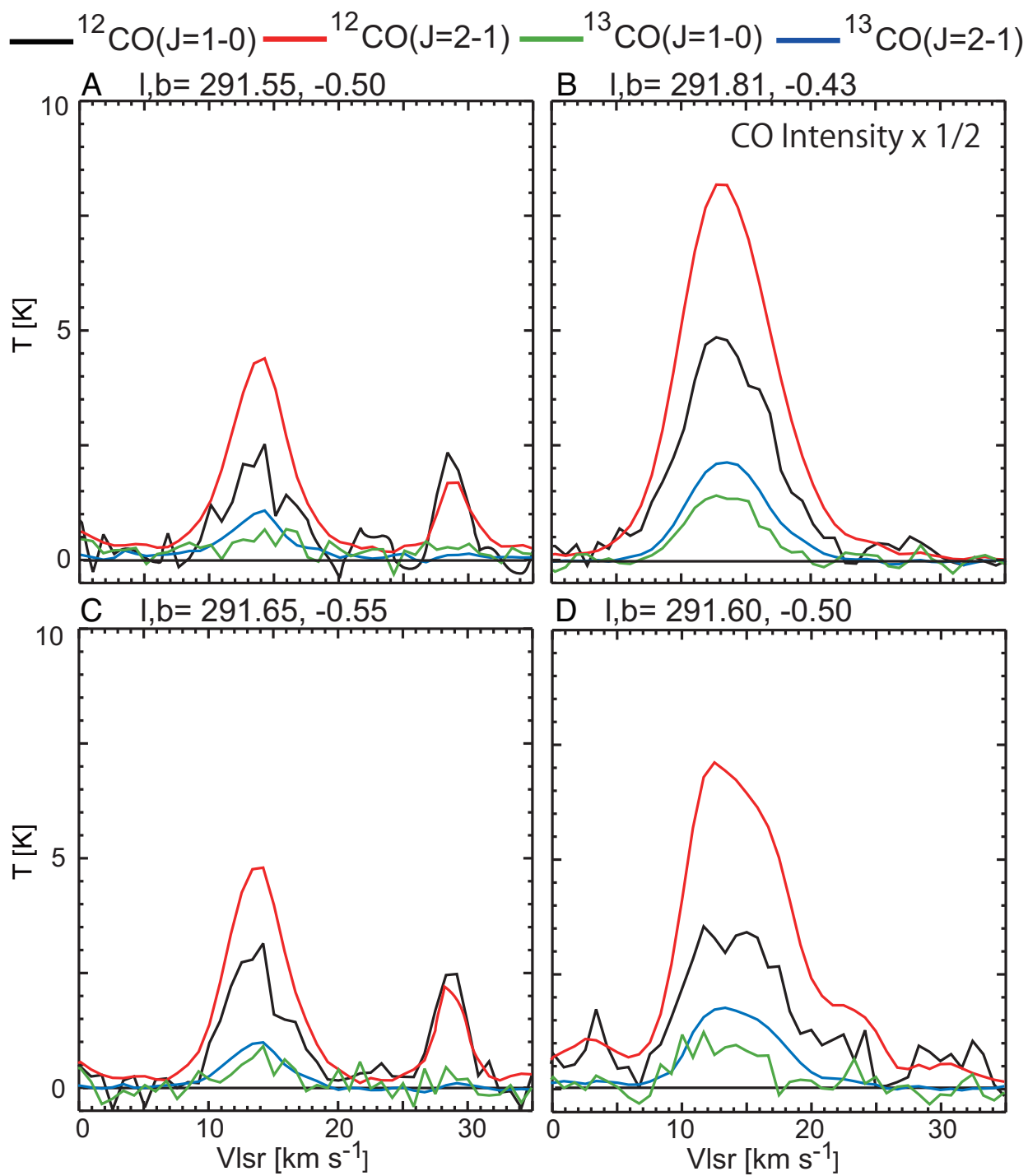}
\caption{CO spectra at the peak positions of the clouds shown in Figure \ref{fig:intensity_ratio}.
CO intensity is plotted at half of its true value for point B.
$^{12}$CO($J$=1--0), $^{12}$CO($J$=2--1), $^{13}$CO($J$=1--0) and $^{13}$CO($J$= 2--1) are plotted in black, red, green and blue, respectively.
All spectra were smoothed to be a beam size of 2{\farcm}6.}
\label{fig:CO_spectra}
\end{figure}

We chose here four positions for a detailed analysis of temperature and density.
They are shown by letters A -- D in Figure \ref{fig:intensity_ratio} and their coordinates are given in Table \ref{tbl:LVG}.
The CO line profiles are shown in Figure \ref{fig:CO_spectra}.
All four lines of $^{12}$CO and $^{13}$CO are detected in the four positions in the blue-shifted cloud, while only the two $^{12}$CO lines are detected in the two positions in the red-shifted cloud. 

\begin{figure}
\epsscale{0.8}
\plotone{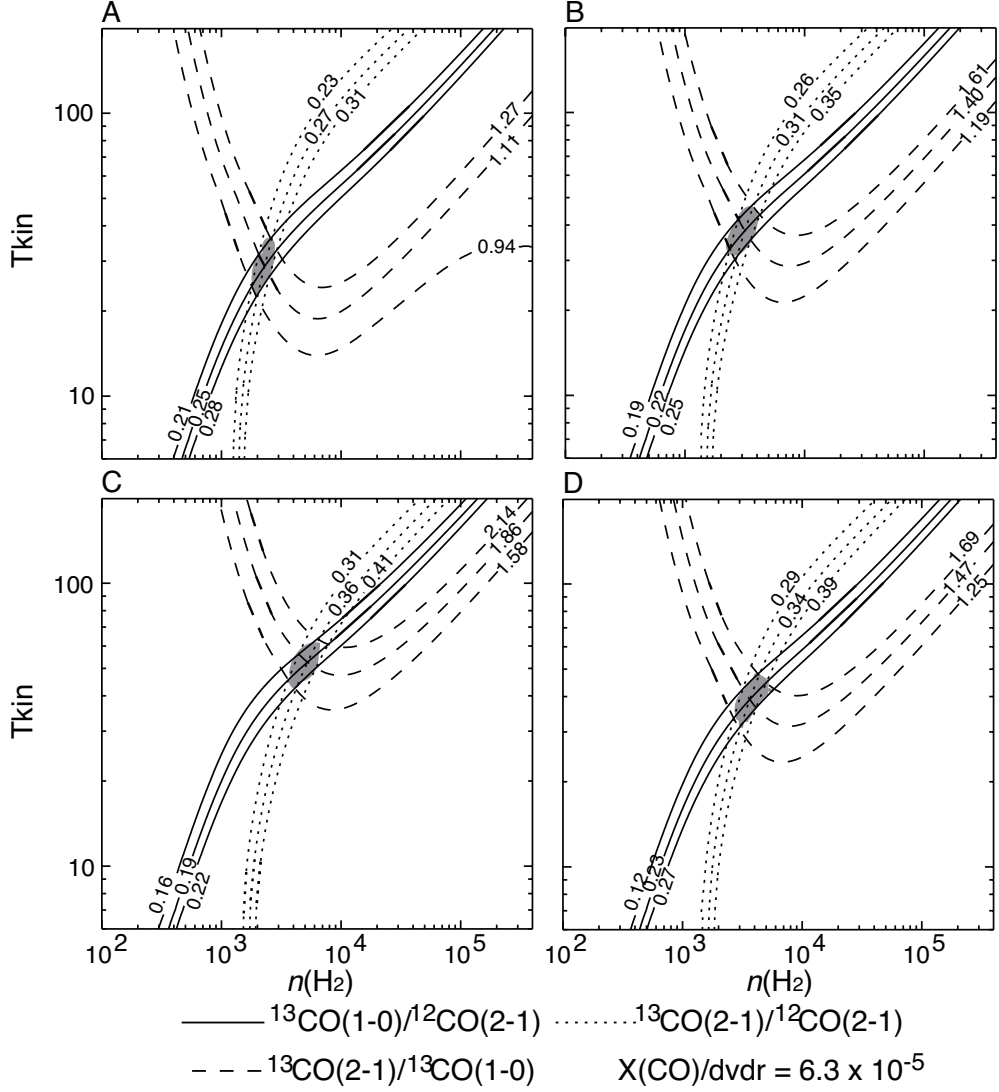}
\caption{LVG results for $X(\mbox{CO})/(dv/dr) = 6.3 \times 10^{-5}$\,(km\,s$^{-1}$\,pc$^{-1}$)$^{-1}$, assuming a distance of 7.0\,kpc, are shown in the density-temperature plane.
Solid, dotted, and dashed lines show $^{13}$CO($J$=1--0)/$^{12}$CO($J$=2--1), $^{13}$CO($J$=2--1)/$^{12}$CO($J$=2--1), and $^{13}$CO($J$=2--1)/$^{13}$CO($J$=1--0) intensity ratios.}
\label{fig:LVG1}
\end{figure}

In order to estimate the kinetic temperature and number density of the molecular clouds, we carried out an LVG analysis \citep{goldreich74}.
The employed model assumes a spherically symmetric cloud where kinetic temperature $T_\mathrm{kin}$, number density $n(\mathrm{H}_{2})$ and the radial velocity gradient $dV/dr$ is taken to be uniform.
We varied  $T_\mathrm{kin}$, and $n(\mathrm{H}_{2})$  within $T_\mathrm{kin} = 6$ -- 500\,K and $n(\mathrm{H}_{2}) =10^{2}$ -- $10^{6}$\,cm$^{-3}$, where we fix $X(\mbox{CO})/(dv/dr) = 6.3 \times 10^{-5}$\,(km\,s$^{-1}$\,pc$^{-1}$)$^{-1}$.
We assume $X(\mbox{CO}) = [^{12}\mbox{CO}]/[\mathrm{H}_2] = 10^{-4}$ \citep[e.g.,][]{1982ApJ...262..590F, 1984ApJS...56..231L} and a velocity gradient of 1.4\,km\,s$^{-1}$\,pc$^{-1}$.
This value of $dv/dr$ was derived by taking the average ratio between the cloud size and velocity width for the four clouds shown in Figure \ref{fig:CO_spectra} (A--D).
For the isotope ratio of $^{12}$C/$^{13}$C, we adopt 75 at the Galactocentric distance of $\sim$9\,kpc \citep{2005ApJ...634.1126M}.
We derived $T_\mathrm{kin}$ and $n(\mathrm{H}_{2})$ in the four positions of the blue-shifted cloud.
We used the line intensity ratios of the $^{12}$CO($J$=2--1), $^{13}$CO($J$=2--1) and $^{13}$CO($J$=1--0) transitions.
The $^{12}$CO($J$=1--0) line was not used here because the line may be optically thick and sample mainly the surface layer the cloud.
Figure \ref{fig:LVG1} shows the results and the derived values are listed in Table \ref{tbl:LVG}.
The temperature is significantly enhanced with respect to quiescent molecular cloud temperatures, falling in the range 30 -- 50\,K in the blue-shifted cloud for a density of 3 -- $5 \times 10^{3}$\,cm$^{-3}$, confirming significant heating by the cluster. 

\begin{figure}
\epsscale{0.6}
\plotone{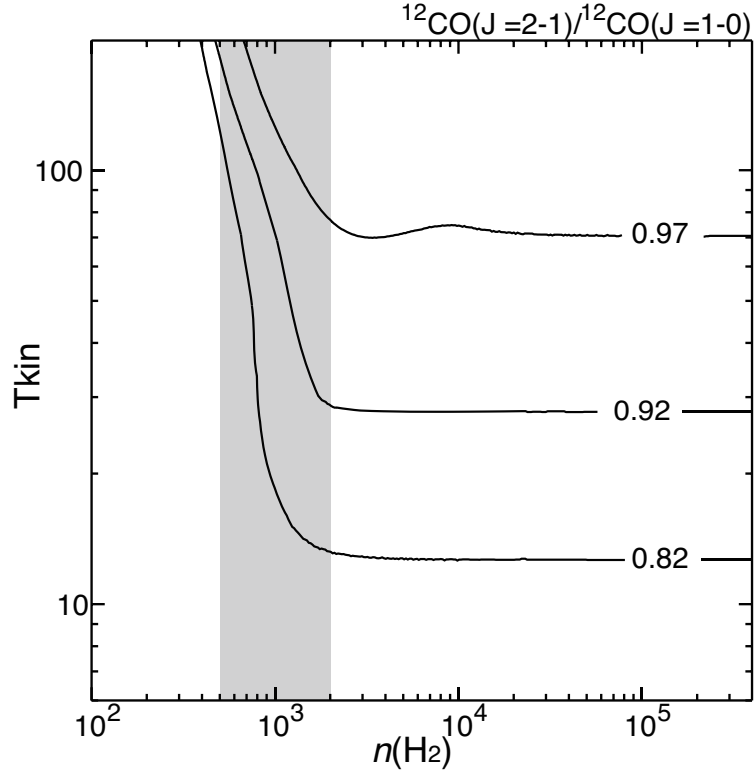}
\caption{LVG results for $X(CO)/(dv/dr) = 6.3 \times 10^{-5}$\,(km\,s$^{-1}$\,pc$^{-1}$)$^{-1}$, assuming a distance of 7.0\,kpc, are shown in the density-temperature plane.
Solid lines show $^{12}$CO($J$=2--1)/$^{12}$CO($J$=1--0) intensity ratios at the position of the red-shifted cloud toward NGC3603.
Gray shows the probale range of number density estimated by the molecular column density and the cloud width.}
\label{fig:LVG2}
\end{figure}

We then estimate the temperature range of the red-shifted cloud from the line intensity ratio of $^{12}$CO($J$=2--1) to $^{12}$CO($J$=1--0).
Figure \ref{fig:LVG2} shows the ratio as a function of density and temperature, where the line intensity is calculated by the LVG approximation.
The molecular column density at positions A and C is estimated from the $^{12}$CO($J$=1--0) integrated intensity by using an empirical X-factor of $2.0 \times 10^{20}$\,cm$^{-2}$\,(K\,km\,s$^{-1}$)$^{-1}$.
By assuming the width of the cloud to be $\sim$1\,pc from the CO($J$=2--1) distribution, we estimate that the density is lower than $\sim 10^{3}$\,cm$^{-3}$ and that the secure lower limit for temperature is estimated to be higher than 20\,K for the $3\sigma$ error limit in Figure \ref{fig:LVG2}.
This indicates that the red-shifted cloud is also heated-up by the cluster NGC3603, the only known strong heat source toward the direction. 

Based on the analysis above, we conclude that the two clouds are located within $\sim$10\,pc of the cluster and adopt 7\,kpc as the distance, the same value as that of the cluster, instead of their kinematic distances of 8 -- 9\,kpc.
For this distance the cloud masses are estimated to be $7.2 \times 10^{4}$\,$M_{\sun}$ and $1.2 \times 10^{4}$\,$M_{\sun}$ for the blue-shifted cloud and the red-shifted cloud, respectively (Table \ref{tbl:cloud_parameters}).

\section{DISCUSSION}\label{sec:discussion}
Previous work has shown that the 13\,km\,s$^{-1}$ molecular cloud is physically linked to the super star cluster NGC3603 as shown by the PDR irradiated by the cluster \citep{2011A&A...525A...8R}.
The present observations confirm this, and also show that there is another extended molecular cloud at 28\,km\,s$^{-1}$ toward NGC3603.
We suggest that the 13\,km\,s$^{-1}$ molecular cloud (blue-shifted cloud) and the 28\,km\,s$^{-1}$ molecular cloud (red-shifted cloud) are both physically associated with NGC3603.
The association is verified by the high temperature of the molecular clouds toward the cluster as derived from an analysis of multi CO transitions, and is consistent with the morphological correlation between CO and NGC3603.
This is a similar situation to the two molecular clouds associated with the super star cluster Westerlund2/RCW49.
For Westerlund2 it is suggested that a collision between the two clouds triggered formation of the cluster, where the relative velocity is ascribed to the original bulk motion of the clouds \citep[Paper I;][]{2010ApJ...709..975O}, and it is possible that a similar collisional process is also working in NGC3603. 

By considering possible projection effects, the observed line-of-sight velocity separation gives a lower limit for the actual relative velocity.
If we assume random cloud motions primarily restricted to the Galactic plane and adopt $\sqrt 2\times 15\sim 20$\,km\,s$^{-1}$ as the relative velocity, the total mass required to gravitationally bind the two clouds is $10^{6}$\,$M_{\sun}$ within 10\,pc of the cluster.
It is one order of magnitude larger than the total mass inside the system $\sim 10^{5}$\,$M_{\sun}$.
We examine an idea that the cloud velocity separation is due to feedback from the cluster or nearby objects.
Supernova remnants (SNRs) may be a possible source of the kinetic energy.
There are two SNRs toward the region, SNR G292.5$-$0.1 \citep{1996A&AS..118..329W} and SNR G292.5$-$0.5 \citep{2001ApJ...554..152C} with a pulsar PSR J1119$-$6127 \citep{2007Ap&SS.308...89G, 2008ApJ...684..532S}.
These are, however, separated from NGC3603 by 0.5\,degrees and are not likely to be affecting the molecular clouds. 
It is also possible that SNRs older than $10^4$\,yr nearby may have affected the cloud motion via SN shock waves in the last Myrs.
If they were influential, one may expect curved velocity field in the two GMCs which may last over Myrs.
The velocity distributions of the present two GMCs however show fairy uniform peak velocity distribution, suggesting such stellar acceleration was not influential.

The velocity difference is not likely due to the stellar-wind acceleration, either.
It is noted that the stellar winds in NGC3603 are affecting an area only within 1\,pc of the cluster \citep{1980ApJ...242..584B,1986MNRAS.219..895C,1990MNRAS.246..712C}, while the \ion{H}{2} region ionized by the cluster is extended within a 10\,pc radius \citep{1986MNRAS.219..895C,1990MNRAS.246..712C}.
The red-shifted cloud is extended far beyond 10\,pc with no large velocity shift in the order of 10\,km s$^{-1}$, indicating that the cloud is not strongly affected by the stellar winds, while a small velocity red-shift of $\sim$ 2\,km\,s$^{-1}$ toward the cluster within 10\,pc in the red-shifted cloud (Figure \ref{fig:pvdiagram}) may possibly be ascribed to the stellar-wind acceleration.
The kinetic energy of this shift is estimated to be $\sim 3 \times 10^{47}$\,erg, and can be supplied by the stellar winds having kinetic energy $\sim 5 \times 10^{51}$\,erg for a 1\,Myr timescale of the highest mass O stars \citep{2008ApJ...675.1319H}, which are most effective in the stellar winds \citep{1995AJ....110.2235D}.
If the velocity separation between the two clouds are mainly due to the stellar winds, the velocity separation of the compact blue-shifted cloud 20\,km\,s$^{-1}$ might be ascribed to such acceleration.
The kinetic energy of the blue-shifted cloud relative to the red-shifted cloud is $\sim 1.6 \times 10^{50}$\,erg for the cloud mass $8.4 \times 10^ {4}$\,$M_{\sun}$ and the relative velocity 20\,km\,s$^{-1}$.
This energy corresponds to several percent of the stellar wind energy, and the acceleration can be possible if the energy requirement alone is considered.
A serious difficulty here, however, is that the blue-shifted cloud shows no such a trend that the velocity separation from the red-shifted cloud becomes large toward the cluster at a pc scale.
The cluster is apparently located on the axis of the cloud elongation in the north to south, and the cloud geometry suggests that most of the cloud cannot be exposed to the winds from cluster.
This is a quite unfavorable configuration for the cloud to be accelerated.
Remembering that the stellar winds are affecting only 1\,pc radius \citep{1980ApJ...242..584B,1986MNRAS.219..895C,1990MNRAS.246..712C}, we conclude it highly unlikely that the blue-shifted cloud as a whole was accelerated by the stellar winds of NGC3603.  

We consider that the association of the two clouds is by chance, and that they were moving independently before their encounter.
The relative velocity 20\,km\,s$^{-1}$ is likely due to random motion of the clouds.
We here present a scenario that a cloud-cloud collision between the two clouds triggered the formation of NGC3603.
The bridging feature discussed in Section \ref{subsec:morphology} suggests that the two clouds are physically interacting; numerical simuations of cloud-cloud collisions find the intermediate velocity features between the two colliding clouds \citep{1992PASJ...44..203H,2010MNRAS.405.1431A, 2012MNRAS.427.1713A}.
In these models two colliding clouds form a compressed layer which is highly turbulent and dense, leading to the formation of dense clumps where high-mass stars are formed.
NGC3603 is the second case of such collision-induced formation of a super star cluster along side Westerlund 2, if the scenario is correct.
In NGC3603, we infer that the collision took place $\sim$1\,Myrs ago as estimated by the ratio of the cloud size 20\,pc and the velocity separation 20\,km\,s$^{-1}$.
This is obviously an order-of-magnitude estimate at best and the value may be different by a factor of $\sim 2$ due to projection effects in space and velocity.
It is however quite unnatural that the time scale can be different by a factor of ten.
The blue-shifted cloud is located on the near side of NGC3603 ($J\!H\!K$ in Figure \ref{fig:blueshifted_cloud} b), which is consistent with a collision that occurred in the past.
A similar argument is given in M20 \citep{2011ApJ...738...46T}.
The blue-shifted cloud has a molecular mass of $7.2\times 10^{4}$\,$M_{\sun}$ and the red-shifted cloud $1.2\times 10^{4}$\,$M_{\sun}$ within a radius of 20\,pc from the cluster.
Both clouds show intensity depressions toward the cluster within a radius of several pc, which includes about ten highest-mass stars having $\sim$100\,$M_{\sun}$ in the cluster core within a radius of 1\,pc \citep{2004AJ....128..765S, 2004AJ....127.1014S, 2008ApJ...675.1319H}.
From these the dissipated cloud mass is roughly estimated to be on the order of $10^{4}$\,$M_{\sun}$ consistent with the mass and the age of the highest-mass members of the cluster core $\sim 10^{4}$\,$M_{\sun}$ and $\sim$1\,Myr \citep{2004AJ....128..765S, 2004AJ....127.1014S, 2008ApJ...675.1319H} for a star formation efficiency of $\sim$10\%.
We suggest that the molecular mass compressed by the collision is around $10^{4}$\,$M_{\sun}$, which corresponds to the mass of a cloud with a molecular column density of $10^{23}$\,cm$^{-2}$ for a radius of 1.5\,pc.
The highest molecular column density observed is $6 \times 10^{22}$\,cm$^{-2}$ toward the northern peak of the blue-shifted cloud, supporting the assumption of such a high column density as the initial condition prior to the collision.

The higher-mass stars above 20\,$M_{\sun}$ have a different slope from that of the lower-mass stars in the mass function in NGC3603 \citep{2008ApJ...675.1319H}.
It is possible that stars of less than 20\,$M_{\sun}$ may have large ages more than several Myrs and may have been already formed prior to the collision.
Recent studies show some low-mass cluster members obviously older than 3\,Myr \citep[e.g.,][]{2012ApJ...750L..44K,2013ApJ...764...73P}, which could be an evidence for the earlier star formation.
The formation of high-mass stars can take place on the order of $10^{5}$\,yrs at a very high mass accretion rate of $10^{-3}$\,$M_{\sun}$\,yr$^{-1}$ as theoretically suggested \citep{2002ASPC..267..267T}.
This high mass accretion rate is sustainable by the increased turbulence in the shocked layer created by the supersonic collision \citep{2010MNRAS.405.1431A, 2013arXiv1301.1016I}.
Another issue is the mass segregation of higher-mass stars toward the cluster \citep[e.g.,][]{2004AJ....127.1014S}; it has been a puzzle how the mass segregation of the higher-mass stars takes place in rich clusters including NGC3603 because gravitational segregation is a long-term process which takes over 10\,Myrs \citep{2007ARA&A..45..481Z, 2008ApJ...675.1319H}.
Cloud-cloud collisions have the potential to cause such mass segregation at the spot of the collisional interaction where molecular density distribution is peaked prior to the collision.

In NGC3603 recent observations indicates that the age spread of the member stars is very short, less than 0.1\,Myrs \citep{2012ApJ...750L..44K}.
This spread is significantly shorter than the sound travel time 1 -- 10\,Myrs for a sound speed of 0.1 -- 1\,km\,s$^{-1}$.
It is notable that the cloud-cloud collision at a high speed of 20\,km\,s$^{-1}$ provides a reasonable explanation for this very short age spread.
In addition, the velocities of the recombination lines and OH absorption lines measured for the nebula surrounding NGC3603 are 8 -- 14\,km\,s$^{-1}$ in $V_\mathrm{LSR}$ \citep[and references therein]{1984ApJ...284..631M}, similar to that of the blue-shifted cloud, having the larger molecular column density than the red shifted cloud.
This is consistent with that the cloud as the parent of the cluster.

The present study has shown that the super star cluster NGC3603 may be the a second super star cluster formed by triggering due to a cloud-cloud collision, along side the super star cluster Westerlund 2.
This suggests that the rare rich star clusters are formed preferentially in the dense interface layer created between two colliding clouds.
As mentioned in Section \ref{sec:introduction}, there are only four super star clusters in the Galaxy that show nebulosity, a hint of the parent cloud(s) of the clusters, but the remainder has no such nebulosity, indicating that their parent clouds are fully dissipated via ionization etc.
It is important to make molecular observations of the other two with associated nebulosity to test if they are also formed by cloud-cloud collisions.

The frequent occurrence of cloud-cloud collisions has been suggested by global numerical simulations of a galactic disk \citep{2009ApJ...700..358T} and recent observations suggest the importance of cloud-cloud collisions in triggering star formation.
Some authors have reported the collision between smaller molecular clouds with 100 -- 1000\,$M_{\sun}$.
For instance, \citet{2011ApJ...738...46T} presented CO($J$=2--1) and CO($J$=1--0) observations of M20 with NANTEN2 and argued that a first generation O-type star \citep{1973ApJ...182L..21W, 1975ApJ...199..647C} was formed by a cloud-cloud collision on the order of 0.5\,Myrs or less.
Also triggered formation is suggested in NGC1333 \citep{1976ApJ...209..466L}, Sgr B2 \citep{1994ApJ...429L..77H, 2000ApJ...535..857S}, W49N \citep{1996MNRAS.281..294B, 2009PASJ...61...39M}, IRAS 0400+5025 \citep{2008ApJ...680..446X}, W51 \citep{2010ApJS..190...58K}, S87, S88B, AFGL5142, AFGL5180 \citep{2010ApJ...719.1813H}, Serpens north \citep{2010A&A...519A..27D}, the stellar cluster L1641-N \citep{2012ApJ...746...25N} and in a further 201 candidates identified from cold IRAS sources \citep{ 2012arXiv1212.0084L}.
We note that some of these observational results only present circumstantial evidence.
The cloud-cloud collision is not a unique interpretation of these observations, because the relatively small velocity separations observed allow the clouds/clumps to be gravitationally bound.
The colliding clouds in the two SSCs, NGC3603 and Westerlund 2, Sgr B2 and M20 have large velocity separations of 8 -- 30\,km\,s$^{-1}$ and they are not gravitationally bound by the observed cloud mass.
The argument on gravitational binding is crucial in order to verify the cloud-cloud collision by excluding gravitationally bound motions.
In order to better understand the role of cloud-cloud collisions in triggering star formation, it is required to have a larger sample of cloud-cloud collisions identified from gravitationally unbound systems.

\section{SUMMARY}\label{sec:summary}

The conclusions of the present work are given as follows;

1) We have found two molecular clouds at around 13\,km\,s$^{-1}$ and 28\,km\,s${-1}$ toward NGC3603 and have shown both are physically associated with NGC3603.
This association is verified by the higher temperature of the molecular gas toward the \ion{H}{2} region as indicated by enhanced CO($J$=2--1/1--0) intensity ratios, and is consistent with the morphological correspondence between the molecular gas and the \ion{H}{2} region and also with a bridging feature in velocity between the two clouds.

2) The velocity separation of the two clouds, 20\,km\,s$^{-1}$, indicates that the two are not gravitationally bound.
Expansion driven by stellar winds also does not provide a good explanation for the cloud motions either.
The physical association of the two clouds therefore must be due to an accidental encounter between them.
We suggest that the two clouds collided with each other $\sim$1\,Myrs ago and this collision triggered the formation of the super star cluster NGC3603 via the strong compression of dense molecular gas in the shocked layer.
	
3) The formation of the super star cluster is very rapid within $\sim$1\,Myr as estimated from the collision timescale and this is consistent with the theory of rapid high-mass star formation by high-mass accretion rate.
The large mass accretion rate is likely sustained by the large turbulence excited by the shock interaction.
This is the second case of formation of a super star cluster by triggering in a cloud-cloud collision along side Westerlund2

\acknowledgments

NANTEN2 is an international collaboration of ten universities, Nagoya University, Osaka Prefecture University, University of Cologne, University of Bonn, Seoul National University, University of Chile, University of New South Wales, Macquarie University, University of Sydney and Zurich Technical University.
The work is financially supported by a Grant-in-Aid for Scientific Research (KAKENHI, Nos.\ 15071203, 21253003, and 20244014) from MEXT (the Ministry of Education, Culture, Sports, Science and Technology of Japan) and JSPS (Japan Soxiety for the Promotion of Science) as well as JSPS core-to-core program (No.\ 17004).
We also acknowledge the support of the Mitsubishi Foundation and the Sumitomo Foundation.
This research was supported by the Grant-in-Aid for Nagoya University Global COE Program, ``Quest for Fundamental Principles in the Universe: from Particles to the Solar System and the Cosmos'', from MEXT.
Also, the work makes use of archive data acquired with \textit{Spitzer Space Telescope} and IRAS data gained with Infrared Processing and Analysis Center (IPAC).
\textit{Spitzer} is controlled by the Jet Propulsion Laboratory, California Institute of Technology under a contract with NASA.

\bibliographystyle{apj}
\bibliography{ohama09bib}
\clearpage







\clearpage

\clearpage

\clearpage

\clearpage

\clearpage

\clearpage

\clearpage

\clearpage

\clearpage

\clearpage

\clearpage

\clearpage

\clearpage

\clearpage
\thispagestyle{empty}
\begin{deluxetable}{lrrrrrrlll}
\rotate
\tablecaption{Properties of the super star clusters in the Galaxy. \label{tbl:ssc_properties}}
\tablewidth{0pt}
\tablehead{\colhead{Name} &\colhead{$l$} &\colhead{$b$} &\colhead{$D$} &\colhead{Age} &\colhead{$\log (M_\mathrm{phot}/M_\sun)$} &\colhead{Radius} &\colhead{IR nebulosities} &\colhead{Reference$^{1}$} &\colhead{Reference$^{2}$} \\
\colhead{ } &\colhead{[degree]} &\colhead{[degree]} &\colhead{[kpc]} &\colhead{[Myr]} & &\colhead{[pc]} &  &  & }
\startdata
Arches &0.12 &0.02 &8.0 &2.0 &4.3 &0.4 &No &1 & \\
Quintuplet &0.16 &$-$0.06 &8.2 &4.0 &4.0 &2.0 &No &2 & \\
RCW38 &268.03 &$-$0.98 &1.7 &$\le$1.0 &--- &0.8 &Yes &3 & \\
Westerlund 2 &284.25 &$-$0.40 &5.4 &2.0 &4.0 &0.8 &Yes &4 &8 \\
Trumpler 14 &287.41 &$-$0.58 &2.6 &2.0 &4.0 &0.5 &No &5  & \\
NGC3603 &291.62 &$-$0.52 &7.0 &2.0 &4.1 &0.7 &Yes &6 &previous work \\
Westerlund 1 &339.55 &$-$0.40 &5.2 &3.5 &4.5 &1.0 &No &7 & \\
$[$DBS2003$]$179 &347.58 &0.19 &7.9 &3.5 &3.8 &1.2 &Yes &4& \\
\enddata
\tablecomments{
Column 1: name of cluster;
Columns 2 and 3: position of cluster;
Column 4: distance;
Column 5: age of cluster;
Column 6: mass of cluster;
Column 7: radius of cluster;
Column 8: comment; 
Column 9: reference$^{1}$ shows paper of clusters; 1, \citet{figer99}; 2, \citet{figer99}; 3, \citet{1987MNRAS.228..721M}; 4, \citet{2009A&A...498L..37P}; 5, \citet{2007yCat..34760199A}; 6, \citet{2008ApJ...675.1319H};  7, \citet{2010A&A...514A..87C, 2005A&A...434..949C}; reference$^{2}$ shows paper of molecular clouds; 8 \citet{2009ApJ...696L.115F, 2010ApJ...709..975O}}
\end{deluxetable}

\clearpage

\begin{deluxetable}{lrrrrrrrr}
\rotate
\tablecaption{Observed properties of the blue-shifted cloud and the red-shifted cloud. \label{tbl:cloud_properties}}
\tablewidth{0pt}
\tablehead{\colhead{Name} &\colhead{$l$} &\colhead{$b$} &\colhead{$W\left[^{12}\mbox{CO}(J=1\mbox{--}0)\right]$} &\colhead{$N_\mathrm{H_2}$} &\colhead{$A_{V}$} &\colhead{$A_{J}$} &\colhead{$A_{H}$} &\colhead{$A_{K}$} \\
\colhead{} &\colhead{[degree]} &\colhead{[degree]} &\colhead{ [K\,km\,s$^{-1}$]} &\colhead{[$10^{21}$\,cm$^{-2}$] } &\colhead{[mag]} &\colhead{[mag]} &\colhead{[mag]} &\colhead{[mag]} }
\startdata
Blue-shifted Cloud North &291.58 &$-$0.42 &150 &30 &31.8 &9.0 &6.0 &3.6 \\
Blue-shifted Cloud South &291.64 &$-$0.55 &80 &16 &17.0 &4.8 &3.2 &1.9 \\
Red-shifted Cloud North &291.56 &$-$0.50 &6 &1.2 &1.3 &0.4 &0.2 &0.1 \\
Red-shifted Cloud South &291.64 &$-$0.58 &12 &2.4 &2.5 &0.7 &0.5 &0.3 \\
\enddata
\tablecomments{
Column 1: name of the cloud.
Columns 2 and 3: peak position of the cloud.
Column 4: integrated intensity of the $^{12}$CO($J$=1--0) emission.
Column 5: molecular column density.
Column 6: visual extinction given by $A_{V}/N_\mathrm{H} = 5.3\times 10^{-22}$\,mag\,cm$^{2}$\,H$^{-1}$ \citep{1978ApJ...224..132B} and $N_\mathrm{H_2}=2N_\mathrm{H}$.
Here we assumed $R_{V}=A_{V}/E(\bv)=3.1$.
Column 7 -- 9 : extinction at $J$ (1.2\,{\micron}), $H$ (1.7\,{\micron}), and $K$ (2.2\,{\micron}) bands, respectively.
Here we adopted an extinction law of \citet{1989ApJ...345..245C}; $A_{J}/A_{V}=0.28$, $A_{H}/A_{V}=0.19$, and $A_{J}/A_{V}=0.11$.}
\end{deluxetable}

\begin{deluxetable}{llrrrrrrr}
\tablecaption{Results of the LVG analysis.\label{tbl:LVG}}
\tablewidth{0pt}
\tablehead{
\colhead{Name} & \colhead{Positon} & \colhead{$l$} & \colhead{$b$} & \colhead{$V_\mathrm{lsr}$} & \colhead{$R_{1}$} & \colhead{$R_{2}$} & \colhead{$T_\mathrm{kin}$} & \colhead{$n(\mbox{H}_2)$} \\
 & & \colhead{[deg]} & \colhead{[deg]} & \colhead{[km\,s$^{-1}$]} & & & \colhead{[K]} & \colhead{[cm$^{-3}$]} }
\startdata
Blue-shifted cloud &A &291.55 &$-$0.50 &12.4--17.7 &1.11 &0.70 &$29^{+9}_{-7}$ &$2.3^{+0.9}_{-0.5} \times 10^{3}$  \\
&B &291.58 &$-$0.43 &12.4--17.7&1.40 &1.26 &28$^{+10}_{-9}$  &$3.2^{+1.8}_{-2.0} \times 10^{3}$  \\
&C &291.31 &$-$0.50 &12.4--17.7&1.86 &1.45 &53$^{+12}_{-11}$ &$5.2^{+4.8}_{-2.2} \times 10^{3}$  \\ 
&D &291.66 &$-$0.58 &12.4--17.7&1.47 &1.81 &39$^{+11}_{-9}$ &$3.6^{+3.4}_{-0.4} \times 10^{3}$ \\
Red-shifted cloud &C &291.60 &$-$0.50 &26.0--30.0&0.92 &--- &$\leq70$$^{\ast}$ &$\geq10^{3}$$^{\dagger}$  \\
&D &291.66 &$-$0.58 &26.0--30.0&0.82 &--- &$\leq 20$$^{\ast}$ &$\geq10^{3}$$^{\dagger}$  \\
\enddata
\tablecomments{Column 1: name of the cloud; 
Column 2: observed points of the cloud; 
Column 3: galactic longitude of the peak position; 
Column 4; galactic latitude of the peak position; 
Column 5; range of velocity integrated; 
Column 6; ratio of  $^{12}$CO($J$=2--1)/$^{12}$CO($J$=1--0); 
Column 7; ratio of $^{13}$CO($J$=2--1)/$^{13}$CO($J$=1--0); 
Column 8: kinetic temperature; 
Column 9: number density of H$_2$.
$\ast$ lower limit
$\dagger$ upper limit}
\end{deluxetable}

\begin{deluxetable}{llrr}
\tablecaption{Physical parameters of the molecular clouds.\label{tbl:cloud_parameters}}
\tablewidth{0pt}
\tablehead{
\colhead{Name} & &\colhead{Mass} &\colhead{$R$} \\
 & &\colhead{[$10^4$\,$M_{\sun}$]} &\colhead{[pc]}}
\startdata
Blue-shifted cloud & &7.2 & \\
&North &5.5 &12.3 \\
&South &1.7 &8.4 \\ 
Red-shifted cloud & &1.2 & \\
&North &0.4 &5.1 \\
&South &0.8 &6.9 \\
\enddata
\tablecomments{
Column 1: name of the cloud;
Column 2: position of the cloud;
Column 3: cloud mass derived using an $X$-factor of $2.0 \times 10^{20}$\,cm$^{-2}$\,(K\,km\,s$^{-1}$)$^{-1}$;
Column 4: radius of the cloud}
\end{deluxetable}

\end{document}